\documentclass[manuscript]{aastex}

\shorttitle{X-ray pulsations from CTA 1}
\shortauthors{Caraveo et al.}

\begin{document}

\title{X-ray pulsations from the radio-quiet gamma-ray pulsar in 
CTA 1\footnote{Based on observations with XMM-Newton, an ESA science
mission with instruments and contributions directly funded by ESA
member states and the USA (NASA).}}

\author{P.A. Caraveo, A. De Luca\altaffilmark{1,2},
  M. Marelli\altaffilmark{3}, G.F. Bignami\altaffilmark{1}}
\affil{INAF - Istituto di Astrofisica Spaziale e Fisica Cosmica,
Via Bassini 15, I-20133 Milano, Italy}
\email{pat@iasf-milano.inaf.it}

\and

\author{P.S. Ray}
\affil{Space Science Division, Naval Research Laboratory, Washington, DC
  20375-5352, USA}

\and

\author{P. M. Saz Parkinson}
\affil{Santa Cruz Institute for Particle Physics, University of California
  Santa Cruz, CA 95064, USA}

\and

\author{G. Kanbach}
\affil{Max-Planck Institut f\"ur Extraterrestrische Physik, 85748 Garching, Germany}

\altaffiltext{1}{Istituto Universitario di Studi Superiori (IUSS) di Pavia,
Viale Lungo Ticino 56, I-27100 Pavia (Italy)}
\altaffiltext{2}{Istituto Nazionale di Fisica Nucleare, Sezione di Pavia, 
Via Bassi 6, I-27100 Pavia (Italy)}
\altaffiltext{3}{Universit\`a degli Studi dell'Insubria, Via Ravasi 2, 21100
  Varese, Italy}

\begin{abstract}
Prompted by the Fermi LAT discovery of a radio-quiet 
gamma-ray 
pulsar inside 
the CTA 1 supernova remnant, we obtained a 130 ks XMM-Newton observation 
to assess the timing behavior of this 
pulsar.  

Exploiting both the unprecedented photon harvest and the contemporary Fermi LAT timing measurements,
a $4.7\sigma$ 
single peak pulsation is detected, making  PSR J0007+7303 
the second example,
after Geminga,
of a radio-quiet gamma-ray pulsar also seen to pulsate 
in X-rays.  
Phase-resolved spectroscopy shows that the off-pulse portion of the light 
curve is dominated by a power-law, non-thermal spectrum,  while the X-ray peak emission appears 
to be mainly of thermal origin, probably from a polar cap heated by
magnetospheric  
return currents, pointing to a hot spot varying throughout the pulsar rotation.
\end{abstract}


\section{Introduction}
The Fermi LAT discovery \citep{abdo08} of a pulsed gamma-ray signal from the
position of the candidate neutron star  (NS) RXJ0007.0+7303 \citep{halpern04}
 inside CTA 1, a 5,000 to 15,000 y old supernova remnant (SNR) at a
distance of 1.4$\pm$0.3 kpc \citep{pineault93}, heralded a new era in pulsar astronomy.
Besides fostering an extraordinary increase in the number of pulsars
detected in gamma rays, the Fermi LAT first pulsar catalogue \citep{abdo10a}
marks the birth of new pulsar sub-families, such as millisecond \citep{abdo09a}
and  the radio-quiet gamma-ray pulsars \citep{abdo09b}. Indeed, with
the recent detection  of 8 
new
objects \citep{sazparkinson10} the gamma-ray 
discovered
pulsars  now 
account for one third
of the pulsating
NSs seen by Fermi \citep{caraveo10,ray10}. 
While the detection of so many radio-quiet objects
points to a gamma-ray beaming covering a solid angle much larger than the
radio one, one wonders if the difference between radio-quiet and radio-loud
gamma-ray pulsars is just a geometrical one.  

X-ray astronomy offers an
independent way to study NSs by assessing their non-thermal magnetospheric
emission together with the temperature and emitting area on their surface.
Previous X-ray studies of CTA 1 central regions unveiled a central
filled SNR (ASCA and ROSAT observations, \citet{Sew95}) and a point source 
(Chandra and a short XMM observations, \citet{slane04,halpern04}), with a jet-like feature, embedded in a
compact nebula. Standard FFT searches on the XMM data failed to detect pulsation, mainly owing
to the source faintness.  
Redoing the exercise using the Fermi LAT timing information yielded unconvincing results, thus prompting the request of a long XMM-Newton observation. 

Searching for the source pulsation was, indeed, the main goal of our 130 ks long
XMM-Newton observation. The unprecedented harvest of X-ray photons, while
unveiling the Fermi LAT periodicity,  allowed also for a new detailed study of
the source spectral shape.

Using  the newly acquired timing and spectral information, we 
compare the  X-gamma behavior of this puzzling radio quiet NS with well known X-ray emitting gamma-ray pulsars.

\section{Observations and data reduction}
\begin{figure}
\centering
\includegraphics[angle=0,scale=.50]{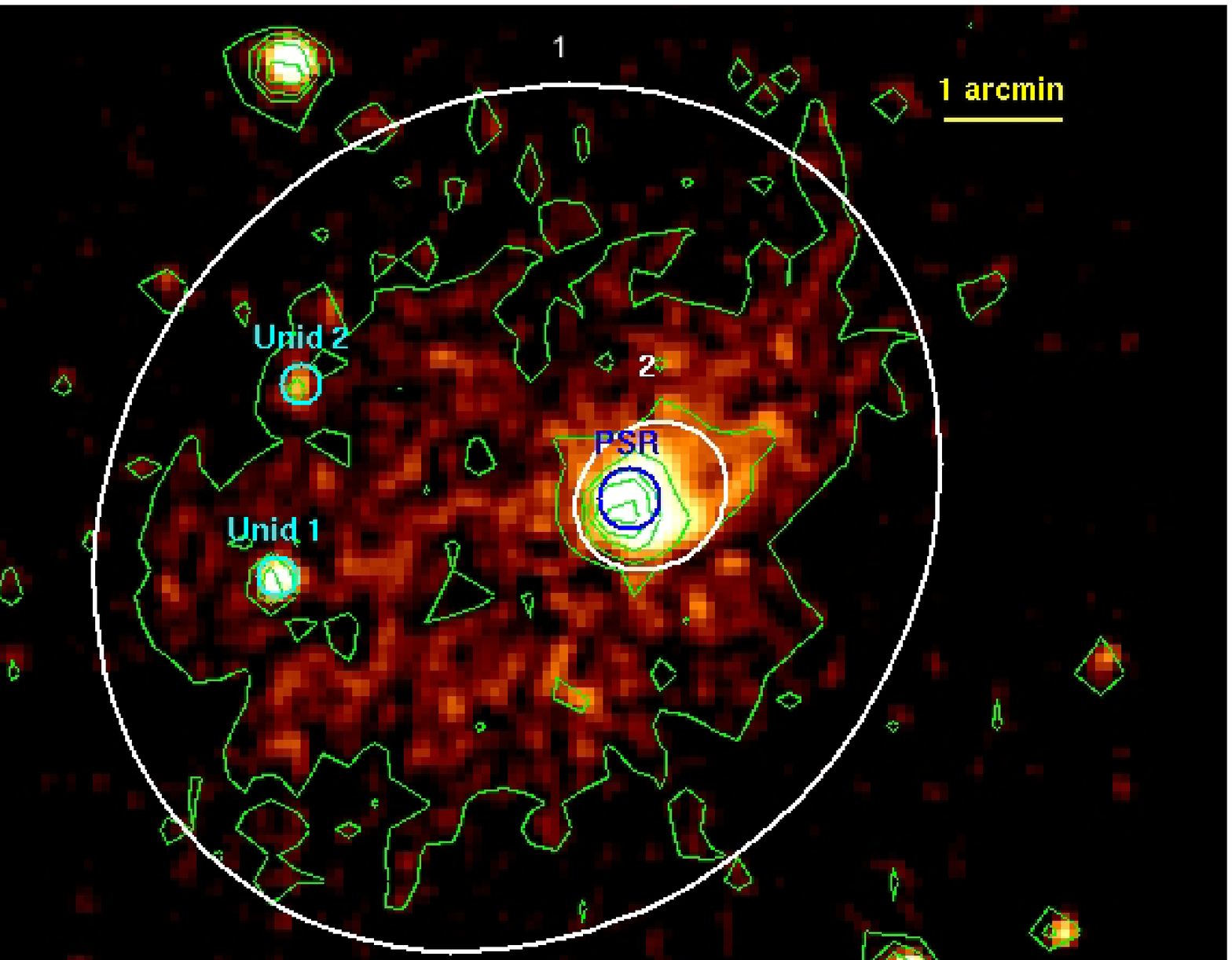}
\caption{0.35-10 keV MOS Imaging. The two MOS exposure-corrected images have been added 
and smoothed with a Gaussian with a Kernel Radius of $13''$. 
The two white ellipses indicate the extended source and the compact PWN.
A $15"$ blue circle indicates the PSR position while two unidentified sources are marked with 
cyan circles.\label{image}}
\end{figure}

Our deep XMM-Newton observation of the CTA 1 system started on 2009, March 7 
at 15:11:10 UT and lasted 130.1 ks. The pn camera \citep{strueder01} 
of the EPIC instrument was operated in Small Window mode 
(time resolution of $\sim5.6$ ms over a $4'\times4'$ field of view), 
while the MOS detectors \citep{turner01} were set in Full frame mode 
(2.6 s time resolution on a 15$'$ radius field of view). 
We used the XMM-Newton Science Analysis Software v8.0. After standard 
data processing (using the {\tt epproc} and {\tt emproc} tasks) and screening 
of high particle background time intervals \citep[following][]{deluca04}, 
the good, dead-time corrected exposure time is 66.5 ks for the PN and 93.5 ks for 
the two MOS. The resulting 0.3-10 keV MOS image 
is shown in Figure ~\ref{image}. 
In order to get a sharp view of the diffuse emission
in the CTA 1 system,
we also used a Chandra/ACIS \citep{garmire03} observation of the field, 
performed on 2003, April 13 (50.8 ks observing time - such dataset was  
included in the investigation by Halpern et al. 2004). 
We retrieved  ``level 2'' data from the Chandra Science Archive and 
used the Chandra Interactive Analysis of Observation (CIAO) software v3.2.

\subsection{Spatial-spectral analysis}
\label{deconv}
The angular resolution of
XMM-Newton  telescopes' is not sufficient to resolve the pulsar (PSR) from the
surrounding pulsar wind nebula (PWN). 
Thus, we used  the spatial-spectral deconvolution method  
developed by \citet{manzali07} 
to disentangle the point source from the diffuse emission,
taking advantage of their different spectra and angular distribution.

\begin{enumerate}
\item for each EPIC instrument, we extracted spectra from three concentric 
regions of increasing radii (0-5$''$, 5$''$-10$''$, 10$''$-15$''$). 
\item based on the well 
known angular dependence of the EPIC Point Spread Function (PSF), we 
estimated the PSR encircled fraction in each region. 
Since the target is on-axis and most of the counts are below 1 keV, 
we used PSF model parameters 
for an energy of 0.7 keV
and null off axis angle. 
\item 
We used Chandra data to compute the PWN encircled 
fraction in each region. To this aim, we simulated 
the PSR point spread function using the ChaRT\footnote{http://cxc.harvard.edu/chart/} 
and MARX\footnote{http://space.mit.edu/CXC/MARX/} software packages, assuming the PSR spectrum 
published by Halpern et al.(2004). We positioned the  
simulated point source at the actual PSR coordinates 
observed by ACIS, and we normalized it in order to match  
the observed peak counts in the ACIS image. Then, 
we subtracted the simulated image from the observed one. 
This yields an image of the diffuse emission 
surrounding the pulsar. However, assuming pure pulsar emission 
at the image peak likely results in underestimating the  
inner (r$\lesssim1.5''$) PWN. To correct the residual image 
for such an effect, we decided to replace counts in the inner  
1.5$''$ with a poissonian distribution having a mean  
value equal to the average number of counts in the  
$1.5''-5''$ surrounding annulus.  
As a last step, following Manzali et al.(2007), we degraded  
the angular resolution to match the EPIC PSF, obtaining a map  
of the PWN surface brightness as seen by EPIC. 

\item we fit a two component (PSR+PWN) model 
to all spectra, freezing the PSR and PWN normalization ratios to the results 
of the previous steps.  Uncertainty in best fit parameters induced by errors
in the encircled fractions is estimated to be
negligible with respect to statistical errors.

\end{enumerate}

Accounting at once for both the PSR and the PWN,
such an approach yields best fit parameters for both 
spectral models. A more detailed description of the method can be found in \citet{manzali07}.
Together with the XMM spectra, we fitted the spectra obtained from the pulsar
(1.5$''$ circle radius) and the nebula (15$''$ circle radius)in the
Chandra observation. We used CIAO 4.1.2 software {\tt acisspec} to
generate the spectrum as well as the response and effective areas.

We focus here into the spectral analysis (step 4). 
Background spectra for each EPIC camera were extracted from source-free
regions within the same chip. Ad-hoc response and effective area 
files were generated using the SAS tasks {\tt rmfgen} and {\tt arfgen}.
Since in our approach encircled energy fractions for the PSR and PWN 
are computed a priori and then used in the spectral analysis, effective
area files are generated with the prescription for extended sources, 
without modelling the PSF distribution of the suorce counts.
Spectra from the three regions were included in a simultaneous fit using
the combination: \\

(interstellar absorption)$\times$(\,$\rho_i$(PSR model)\,+\,$\epsilon_i$(PWN model)\,) \\

where $\rho_i$ and $\epsilon_i$
are the PSR and PWN encircled fractions within the $i^{th}$ extraction region.

The interstellar absorption coefficient does not depend
on $i$. For the PWN, we used a power law model. Although
the PWN spectrum is expected to vary as a function of
the position, 
the relatively small photon statistic
prompted us to fit
a single photon index $\Gamma_{PWN}$ 
to all regions. 
Of course, PSR parameters do not vary in the different annuli.
For the PSR emission, we tried  a simple power law, 
the combination of a power law and a blackbody as well as
the combination of a power law and a magnetized neutron star atmosphere
model ({\em nsa} in XSpec - assuming a 1.4 M$_{\odot}$ neutron star
with a radius of 13 km and a surface magnetic field of $10^{13}$
G).

All the models for the PSR emission yield statistically acceptable fits
(power law: $\chi^2_{\nu}$ =91.5, 124 d.o.f.; 
blackbody+power law: $\chi^2_{\nu}$=85.8, 121 d.o.f.;
{\em nsa}+power law: $\chi^2_{\nu}$=86.8, 121 d.o.f.). 

The resulting parameters are summarized in Table 1. 
As expected, the {\em nsa} model yields smaller temperatures and
larger emitting areas than the blackbody model.

Using the best fit blackbody+power law model, within a 15$''$ 
circle in 0.3-10 keV, we estimate that 
47\%  of the pn counts come from the PSR, 
32\% from the PWN and 21\% are background (instrumental as well as cosmic). 

In order to discriminate between the 
purely non-thermal and the composite (thermal + non-thermal)
description of the
pulsar emission, the high resolution,
temporal information provided by the pn instrument
is crucial.

\subsection{Timing Analysis}
4989 pn events in the 0.15-10 keV energy range were extracted from a 15$''$ circle, 
centered on the gamma-ray pulsar. PATTERN selection was performed as by 
\citet{pellizzoni08}. X-ray photons' times of arrival were
barycentered 
according to the PSR Chandra position (RA 00:07:01.56, Dec 73:03:08.3) and
then folded according to an accurate Fermi-LAT timing solution \citep{abdo10a}
that overlap our XMM dataset (the pulsar period at the start of our XMM-Newton
observation is P=0.3158714977(3) s).
Such exercise
was repeated selecting photons in different energy ranges.

\begin{figure}
\centering
\includegraphics[angle=0,scale=.50]{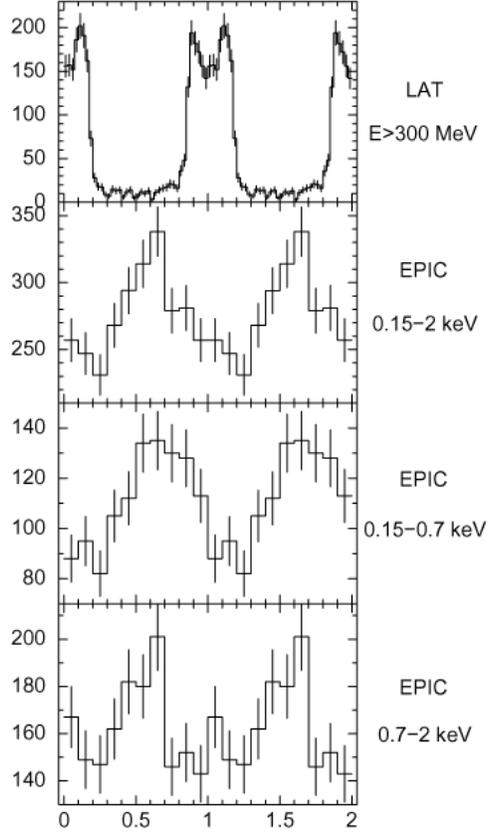}
\caption{EPIC/pn folded light curves in different energy ranges using
photons within a $15"$
radius from the Chandra position. X-ray photons' phases were computed according 
to an accurate Fermi-LAT ephemeris overlapping
with the XMM dataset: the pulsar period at the start of the XMM observation is P=0.3158714977(3) s
and the $\dot{P}$ contribution was taken in account.
PATTERN 0 events have been selected in the 
0.15-0.35 keV energy range while PATTERN $\leq$4 have been used in the 0.35-2 keV range.
The upper panel shows the LAT light curve
of the CTA 1 pulsar from \citet{abdo10a} to which the XMM
light curves have been aligned in phase.\label{lc}}
\end{figure}

A 4.7$\sigma$ pulsation is seen in the 0.15-2 keV energy range 
(null hypothesis probability of 
$1.1\times10^{-6}$, according to a $\chi^2$ test), 
characterized by a single peak, 
which is out of phase with respect to the gamma-ray emission.  Light curves 
computed for different energy ranges are shown in Figure ~\ref{lc}, 
phase aligned with the gamma-ray one.   

Results from the previous section make it possible to compute a net
(background and PWN - subtracted) pulsed fraction of $85\pm15\%$  
in the 0.15-0.75 keV  energy range. No pulsation is seen in the 2-10 keV
energy range. Assuming a sinusoidal pulse profile, we evaluated
a $3\sigma$ upper limit of 57\% on the net pulsed fraction. 
Such a difference in the overall source pulsation as a function of 
the photon energy does not support the single-component model for the PSR
emission, pointing to the composite thermal plus power law model.

\subsection{Phase-resolved spectroscopy}
Phase-resolved spectral analysis  was performed by  selecting on- and
off-pulse portions of the light curve. We selected pn events 
(15$''$ extraction radius, PATTERN 0) from the phase intervals
corresponding 
to the peak and to the minimum of the folded light curve.  These spectra  are 
plotted in Figure \ref{phaseres} where they are seen to differ only at 
low energy,  
while they appear superimposed for E$>1.2$ keV.  We adopted the  
best fit model computed in sect.~\ref{deconv}, featuring
the composite 
black-body + power-law spectrum for the PSR, 
as a template to describe the phase-resolved spectra.
Following \citet{caraveo04} and \citet{deluca05}, we fixed
all spectral parameters (including the PWN component)
at their phase-averaged best fit values and we used the 
the normalizations of the PSR thermal and non-thermal components to describe
the pulse-phase modulation.
The spectral variation may be well
described as a simple modulation of the emitting radius of the 
thermal 
component, keeping the power law component fixed
($\chi^2=26.2$, 31 d.o.f. for the blackbody+power law model; $\chi^2=25.9$, 31 d.o.f. for the {\em nsa}+power law model). 
Using the blackbody model, the emitting radius of the thermal component 
varies from $242_{-242}^{+111}$ m to $600_{-75}^{+68}$ m as a  function of the star rotation phase. 
Using the {\em nsa} description, the emitting radius varies from
$560_{-560}^{+720}$ m to $4380_{-1150}^{+740}$ m. 
Quoted values of the emitting radii are as measured from a distant observer.
Such a variation could easily account for the totality of the X-ray pulsation. 

Fitting the on-off spectra using a single power law component does not yield
acceptable 
results ($\chi^2=48.9$, 31 d.o.f.),
while the paucity of the counts does not allow to test a model where both
thermal and 
non-thermal components are allowed to vary.

\subsection{Extended emission spectral analysis}
Diffuse emission, already
discovered by ROSAT and ASCA \citep{halpern04,slane04},
pervades the entire EPIC/MOS field
of view. A thorough analysis of such emission,
requiring ad-hoc background subtraction/modelling techniques, 
is beyond the scope of this paper, focused the pulsar phenomenology.
For completeness, we include a simple study of the inner and brighter portion
of the diffuse emission (within $\sim150''$ from the 
pulsar). Using the brightness profile along different radial 
directions, we selected two elliptical regions (ellipse 1 and 2,
see Figure 1) -- excluding the inner 15$''$ radius circle --
and we extracted the corresponding spectra from the MOS data. Background spectra
were extracted from a region outside ellipse 2.
Since ellipse 1 lies within the pn field of view, we extracted also a
pn spectrum for such a region. Owing to the dimension of the pn field of view,
the pn background spectrum was extracted 
from a region within ellipse 2. However, the 
difference in surface brightness between the two ellipses is large enough to
induce a negligible distortion to the pn ellipse 1 spectrum.
The extended emission is well described by a power law spectrum
with an index of $1.59\pm0.18$ in the inner portion (ellipse 1, 
$\chi^2$=123.9, 90 dof) and of $1.80\pm0.09$ in the outer portion 
(ellipse 2, $\chi^2$=189.3, 137 dof). The observed flux is of
1.60$\pm$0.09 $\times10^{-13}$ erg cm$^{-2}$ s$^{-1}$ and of 1.98$\pm$0.06 $\times10^{-12}$ erg cm$^{-2}$ s$^{-1}$
for the inner and outer portions, respectively.
Owing to smaller
collecting area as well as larger background per unit solid angle, Chandra/ACIS data yield consistent,although less constrained, results for such an extended 
emission.

\section{Discussion and Conclusions}

After Geminga, PSR J0007+7303 now becomes the second example of a
radio-quiet gamma-ray pulsar also seen to pulsate in X-rays. 

Our deep XMM-Newton observation characterizes the system emission
as follows.

The PWN X-ray spectrum at the position of the pulsar can be described by a power law 
with index $\Gamma_{PWN}=1.5\pm0.3$. Diffuse, non-thermal emission 
with a decreasing surface brightness is seen across the EPIC field of view,
with a photon index steepening as a function of the distance from the pulsar
($\Gamma=1.80\pm0.09$ at $1\div2.5$ arcmin distance).

The X-ray spectrum of the pulsar is a combination of thermal radiation
superimposed to a non-thermal power law component. Thermal emission
comes from a ``hot spot'', larger than the polar cap computed for a dipole model for 
PSR J0007+7303 (about 100 m radius), but significantly smaller than the entire surface 
of any reasonable NS. This is true both using the blackbody model
and the neutron star atmosphere model (see Table 3), although the latter yields
a larger emitting area and a lower temperature. Indeed, we warn that 
for some thermally emitting PSR the use of the {\em nsa} model turned out to be 
problematic (see e.g. the case of PSR B0656+14 discussed by De Luca et al. 2005).
Thus, {\em nsa} results should be taken with caution.
At variance with the majority of X-ray emitting isolated  
pulsars (e.g. \citet{kaspi06}), no thermal component from the whole NS surface 
is discernible from  the XMM spectrum. 
Using the blackbody model, the $3\sigma$ upper limit  on the temperature of a 
10 km radius NS is $5.3\times10^5$ K and $4.8\times10^5$ K using the {\em NSA} model.
This makes PSR J0007+7303 by far 
the coldest NS  for its age interval,  
suggesting a rapid cooling for this young gamma-ray pulsar.   

The detection of X-ray pulsation makes it possible to directly compare the PSR J0007+7303  
multiwavelength phenomenology with that of other prototypical pulsars.

With a rotation energy loss of 4.52 $\times 10^{35} erg s^{-1}$ and a kinematic age of 13 ky, PSRJ0007+7303
is 50 times younger and 10 times more energetic than Geminga, for many years
the only known radio-quiet gamma-ray pulsar (for a review, see Bignami \&
Caraveo, 1996).  While PSR J0007+7303 is a relatively young,  Vela-like pulsar, its
rotational energy loss is intermediate between Geminga and Vela. Thus, it makes sense to
compare PSR J0007+7303 with Vela and Geminga, two NSs with a well
established  multiwavelength phenomenology \citep{San02,manzali07,caraveo04,deluca05,jackson05}.

Starting from the source flux values, if we consider the ratio between the
gamma-ray and non-thermal X-ray fluxes, we find a value of $(5.6 \pm 1.1) \times10^3$ for
CTA 1, to be compared with $(6.8 \pm 0.4) \times 10^3$ for Geminga  and
$(1.3\pm0.3) \times 10^3$ for
Vela. Thus, the young and energetic gamma-ray pulsar PSR J0007+7303 is
somewhat under-luminous in X-rays, joining  Geminga  and PSR J1836+5925, another
radio-quiet pulsar also known as Next Geminga \citep{halpern07,abdo10b}.

Turning now to the phase resolved spectral analysis, we note that the peak
emission of the newly measured  single-peak X-ray light curve can be
ascribed to a hot spot,  apparently varying
throughout the pulsar rotation. Although the hot spot dimension seems too
big to be reconciled with the NS polar cap, but too small to account for
the entire NS surface, the varying thermal contribution  is indeed
reminiscent of of the behavior of middle-aged pulsars such as Geminga and PSR
B1055-52 \citep{deluca05} .


Although PSR 0007 is the second radio quiet NS seen to pulsate in X-ray, we
note that the X-to-gamma-ray observational sequence, which was successfully
applied with Geminga, has been reversed.
Here, for the first time, the basic pulsar properties (P, $\dot{P}$) were
discovered through the gamma-ray emission which made it possible the
detection of  the source X-ray pulsations.

While X-ray
 observations of gamma-ray pulsars remain crucial to probe the source
physics, the superb sensitivity of the Fermi LAT has made the discovery of
periodicities far more fruitful in the gamma-ray band, as shown by
\citet{abdo09b} and \citet{sazparkinson10}.

\acknowledgments

This work was supported by contracts ASI-INAF I/088/06/0 and NASA NIPR NNG10PL01I30.

{\it Facilities:} \facility{XMM (EPIC)}, \facility{CXO (ACIS)}.

\clearpage

\begin{figure}
\centering
\includegraphics[angle=0,scale=.40]{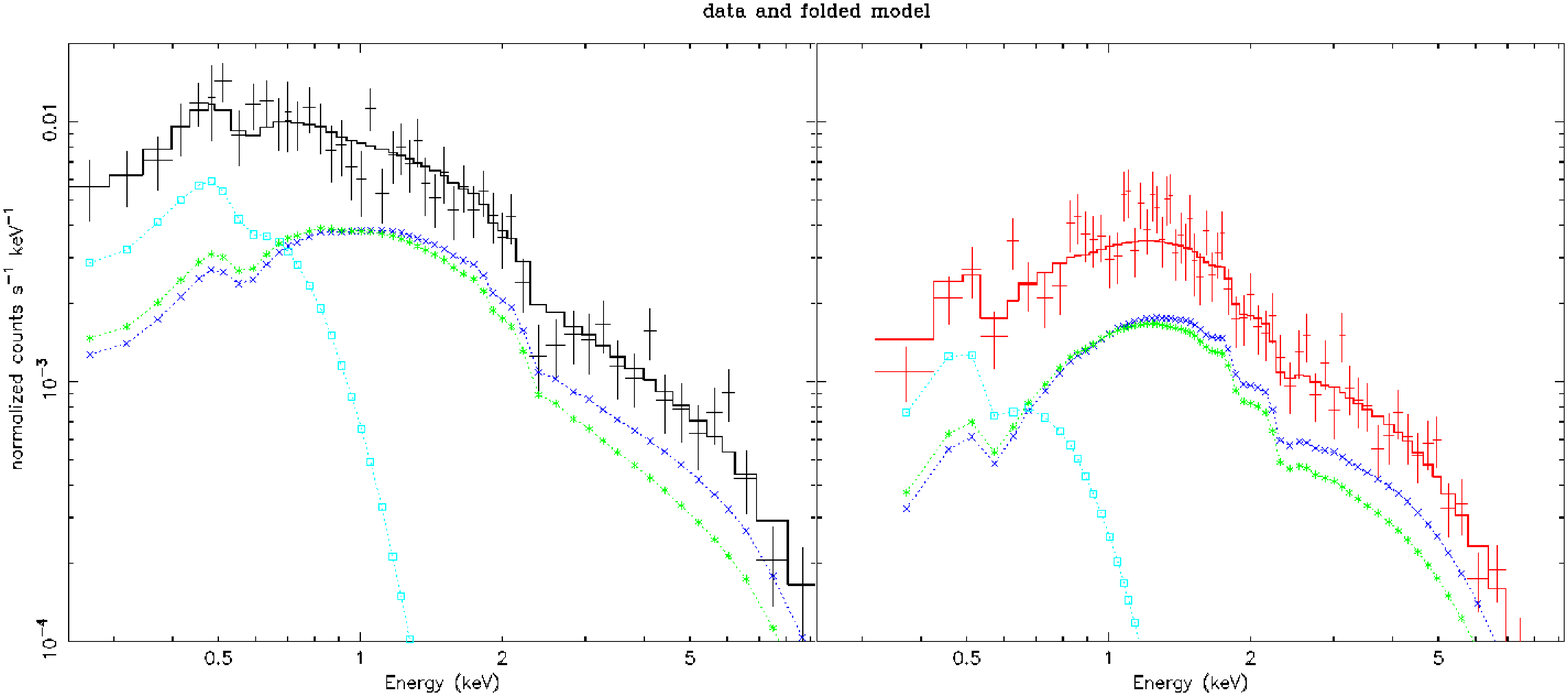}
\caption{PN and MOS spectra of PSR J0007+7303.
PATTERN 0 PN events  and PATTERN $\leq$12 MOS events 
have been selected among photons within 15$''$ from the target position.
The spectra are rebinned in order to have at least 25 counts
per bin and no more than 3 spectral bins per energy resolution
interval.
The black and the red curves show respectively the PN and MOS data and spectral 
fits.  
Cyan square-marked curve shows the blackbody component, 
while pulsar power law is shown with blue curves and PWN one with green asterisks.
\label{specresults}}
\end{figure}

\begin{figure}
\centering
\includegraphics[angle=0,scale=.50]{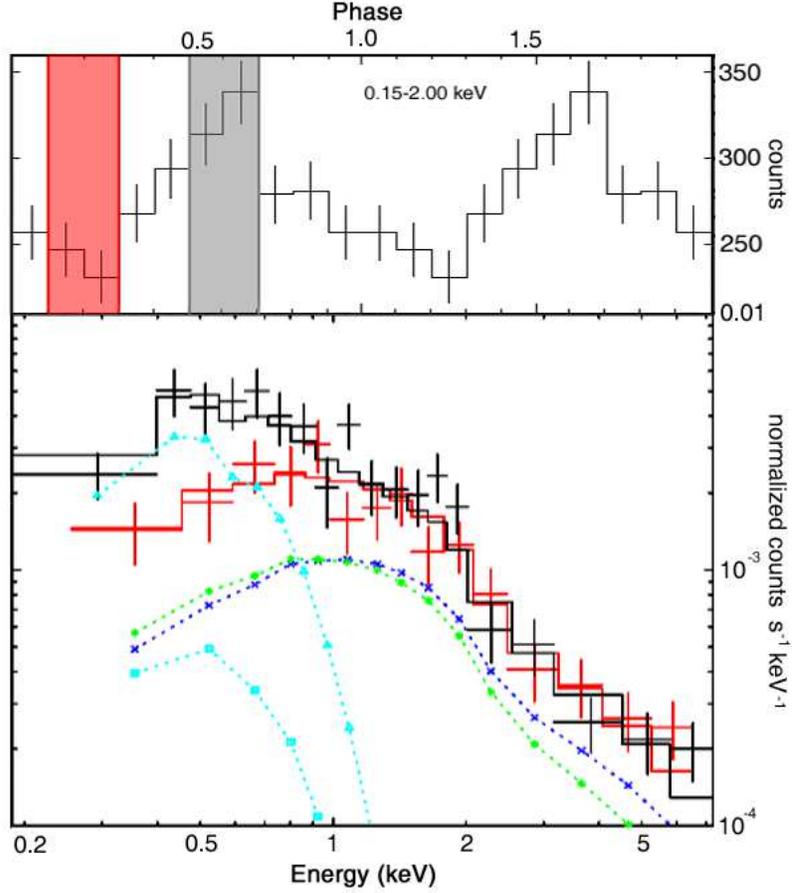}
\caption{
Upper panel: X-ray folded light curve 
of PSR J0007+7303 (0.15-2 keV).
Lower panel: X-ray spectra relative to the phase intervals shaded 
in gray (on-pulse, black line) and red (off-pulse, red line). 
Both spectra were fitted with a 3-component model to account for 
the NS thermal emission (cyan symbols) and power law (blue dotted line) 
as well as the PWN power law (green dotted line). With the power law 
contributions unchanged in the two spectra, a significant thermal component 
is present only in the on-pulse spectrum (cyan triangles) 
while it appears suppressed in the off-pulse one (cyan squares).
\label{phaseres}}
\end{figure}

\clearpage

\begin{table}
\begin{center}
\caption{CTA 1 Pulsar and nebula spectra.\label{tab-1}}
\begin{tabular}{l|rrr|r|r}
\tableline\tableline
 &Pulsar&J0007+7303& &Inner PWN&Outer PWN\\
Parameter & PL & PL+BB & PL+{\em NSA} & & \\
\tableline
 N$_{H}$($10^{21}$)                                & 0.63$_{-0.23}^{+0.25}$ & 1.66$_{-0.76}^{+0.89}$ & 1.53$_{-0.73}^{+1.19}$ & 1.67$_{-1.22}^{+1.38}$ & 1.83$_{-0.27}^{+0.30}$\\
 $\Gamma_{PWN}$                                     & 1.25$_{-0.15}^{+0.17}$ & 1.53$_{-0.27}^{+0.33}$ & 1.49$_{-0.24}^{+0.32}$& 1.59$\pm$0.18 & 1.80$\pm$0.09\\
 $\Gamma_{PSR}$                                     & 1.36$_{-0.14}^{+0.16}$ & 1.30$\pm$0.18 & 1.25$_{-0.19}^{+0.20}$& - & -\\
 kT(keV)                                          & - & 0.102$_{-0.018}^{+0.032}$ & $0.054_{-0.016}^{+0.025}$ & - & -\\
 r$_{1.4kpc}$(km)                                  & - & $0.64_{-0.20}^{+0.88}$ & $4.92_{-4.68}^{+1.81}$& - & -\\
 $\chi^{2}$                                         & 91.56 & 85.81 & 86.82 & 123.89 & 189.30\\
 d.o.f.                                            & 124 & 121 & 121 & 90 & 137\\
Total Flux$_{0.3-10 keV}$ $^a$  & 12.00$\pm$0.10 & 13.90$\pm$0.36 &
14.01$\pm$0.41 & 16.0$\pm$0.9 & 198$\pm$6\\
Total Flux$_{2-10 keV}$ $^a$ & 8.83$_{-0.28}^{+0.37}$ & 8.69$_{-0.86}^{+0.97}$
& 8.74$_{-0.91}^{+1.02}$ & 10.1$\pm$0.6 & 105$\pm$5\\
PSR Flux$_{0.3-10 keV}$ $^a$   & 6.54$\pm$0.53 & 8.41$\pm$0.98 & 8.41$\pm$1.00
& - & -\\
PSR Flux$_{2-10 keV}$ $^a$   & 3.98$\pm$0.72 & 4.30$_{-0.61}^{+1.62}$ &
4.32$_{-0.65}^{+1.67}$ & - & -\\
Thermal Flux $^a$ & - & 1.55$\pm$1.01 & 1.68$\pm$1.11 & - & -\\
\tableline
\end{tabular}
\tablecomments{X-ray spectrum of the pulsar and the nebula.  
Inner and Outer PWN correspond to emission from ellipse 1 and
ellipse 2, respectively (see taxt and Figure 1).  
For the pulsar we provide the power law, 
the blackbody + power law and the magnetized neutron star
atmosphere model ({\em nsa}) + power law spectral fits. Temperatures
and emitting radii are as measured from a distant observer.
For the pulsar and inner nebula we used Chandra, MOS and PN data,
while for the outer PWN spectra we used
only data from MOS1+2 instruments owing to the small FOV of PN.\\ $^a$ Fluxes are in units of $10^{-14}$erg/cm$^2$s}
\end{center}
\end{table}


\begin{thebibliography}{}
\bibitem[Abdo et al.(2008)]{abdo08} Abdo, A., A., et al., 2008, Science 322, 1218
\bibitem[Abdo et al.(2009a)]{abdo09a} Abdo, A., A., et al., 2009a, Science 325, 848
\bibitem[Abdo et al.(2009b)]{abdo09b} Abdo, A., A., et al., 2009b, Science 325, 840
\bibitem[Abdo et al.(2010a)]{abdo10a} Abdo, A., A., et al., 2010a, ApJS 187, 460
\bibitem[Abdo et al.(2010b)]{abdo10b} Abdo, A., A., et al., 2010b, ApJ 712, 1209
\bibitem[Bignami \& Caraveo(1996)]{bignami96} Bignami, G.F. \& Caraveo, P.A.,
  ARA\&A 34, 331
\bibitem[Caraveo(2010)]{caraveo10} Caraveo, P.A., 2010, in High Time Resolution Astrophysics 
(HTRA) IV, Agios Nikolaos, Crete (Greece), May 5 - 7, 2010, arXiv:1009.2421 
\bibitem[Caraveo et al.(2004)]{caraveo04} Caraveo, P.A., et al., 2004, Science
  305, 376
\bibitem[De Luca et al.(2005)]{deluca05} De Luca, A., et al., 2005, ApJ 623,
  1051
\bibitem[De Luca \& Molendi(2004)]{deluca04} De Luca, A. \& Molendi, S., 2004,
  A\&A 419, 837
\bibitem[Garmire et al.(2003)]{garmire03} Garmire, G., et al., 2003, SPIE
  4851, 28
\bibitem[Halpern et al.(2004)]{halpern04} Halpern, J.P., et al., 2004, ApJ 612, 398
\bibitem[Halpern et al.(2007)]{halpern07} Halpern, J.P., et al., 2007, ApJ 668, 1154
\bibitem[Jackson et al.(2005)]{jackson05} Jackson, M.S., et al., 2005, ApJ 633, 1114
\bibitem[Kaspi et al.(2006)]{kaspi06} Kaspi, V., et al., 2006, in ``Compact Stellar X-ray
  Sources'', eds. Lewin, W. and van der Klis, M., Cambridge University Press,
  p. 279 
\bibitem[Manzali et al.(2007)]{manzali07} Manzali, A., et al., 2007, ApJ 669,
  470
\bibitem[Pellizzoni et al.(2008)]{pellizzoni08} Pellizzoni, A., et al., 2008,
  ApJ 679, 664
\bibitem[Pineault et al.(1993)]{pineault93} Pineault, S., et al., 1993, AJ
  105, 1060 
\bibitem[Ray \& Saz-Parkinson(2010)]{ray10} Ray, P.S. \& Saz-Parkinson, P., 2010, 
Proceedings of ICREA Workshop ``The High-Energy Emission from Pulsars and their Systems'',
arXiv:1007.2183
\bibitem[Sanwal et al. (2002)]{San02} Sanwal, D. et al. 2002, ASPC, 271, 353
\bibitem[Saz Parkinson et al.(2010)]{sazparkinson10} Saz Parkinson, P., et al., 2010, 
submitted to ApJ, arXiv:1006.2134
\bibitem[Seward(1995)]{Sew95} Seward, F. et al. 1995, \aap, 453, 284
\bibitem[Slane et al.(2004)]{slane04} Slane, P., et al., 2004, ApJ 601, 1045 
\bibitem[Strueder et al.(2001)]{strueder01} Strueder, L., et al., 2001, A\&A
  365, L18
\bibitem[Turner et al.(2001)]{turner01} Turner, M., et al., 2001, A\&A 365, L27
\end{thebibliography}
\end{document}